\documentclass[12pt]{iopart}
\usepackage{amssymb}
\usepackage{txfonts}
\usepackage{bbm}
\usepackage{graphicx}
\usepackage{appendix}
\usepackage{epsf}
\usepackage{epstopdf}

\begin{document}

\title{Impurity effect as a probe for order parameter symmetry in iron-based superconductors}

\author
{Tao Zhou,${^1}$ Xiang Hu,${^{1}}$ Jian-Xin Zhu,${^2}$ and C. S.
Ting$^{1}$}

\address{$^{1}$Texas Center for Superconductivity and Department of
Physics, University of Houston, Houston, Texas 77204\\
$^{2}$ Theoretical Division, Los Alamos National Laboratory, Los
Alamos, New Mexico 87545} \ead{tzhou2@uh.edu}
\date{\today}

\begin{abstract}

The pointlike impurity scattering is studied for a superconductor
with different pairing symmetries based on a minimal two band model
proposed by S. Raghu $et$ $al.$ [Phys. Rev. B {\bf 77}, R220503
(2008)], and the self-consistent Bogoliubov-de Gennes equations. Two
intra-gap resonance peaks are found at or near the nonmagnetic
impurity site for a typical $s_{x^2y^2}$ symmetry and strong
scattering potentials. Such intra-gap resonance peaks are absent for
the sign unchanged $s$-wave and $d_{x^2-y^2}$-wave order parameter.
Thus we propose that a non-magnetic impurity with strong scattering
potentials can be used to probe the pairing symmetry in iron-based
superconductors.
\end{abstract}

\maketitle

\section{Introduction}

 The new family of superconducting materials which
contain Fe-As layers has attracted much attention since their
discovery~\cite{kam}. The Fe-As layers are believed to be the
conducting planes. The band calculations have shown that the Fermi
surface includes two hole pockets centered at $\Gamma$ points and
two electron pockets centered at $M$ points,
respectively~\cite{sin,ish,nak}, which is confirmed by
angle-resolved photoemission spectroscopy (ARPES)
experiments~\cite{lu,din,kon,naka,evtu,cliu}. A number of
experiments have indicated that this family of materials is not
conventional~\cite{sad}. Therefore, probing the pairing symmetry is
one of the most important issues since it can provide us the
information of the pairing mechanism.

So far, the pairing symmetry in iron-based superconductors is still
controversial. One popular proposal for the pairing symmetry is the
$s_{\pm}$-wave gap~\cite{maz,chu,zjyao,shun,wan,cve}, namely, the
superconducting gap is extended $s$-wave and has opposite sign for
the hole and electron pockets. Experimentally, the $s$-wave gap
symmetry is supported by penetration depth measurement~\cite{has},
ARPES~\cite{din,kon,naka,evtu,cliu} and specific heat
measurement~\cite{mu}. In particular, the gap magnitudes at
different Fermi surface pockets measured by ARPES experiments agree
well with $s_{x^2y^2}=\cos k_x \cos k_y$ form (here $k_x$ and $k_y$
represent the momentum in the unfolded Brillouin zone)~\cite{naka}.
However, although ARPES experiment is a powerful tool to measure the
gap magnitude directly, it cannot determine the phase of the
superconducting order parameter. Detecting the phase of order
parameter is an essential and important step to map out the gap
symmetry. Recently several theoretical works propose different ways
to measure the sign change of the gap~\cite{tsa,gha,par,wu,zha}.
While for multiband materials, probing the phase of the order
parameter is still a quite challenging task.

The impurity effect has been an important part in the studies of the
superconductivity~\cite{bal}. The effect of the impurity scattering
is sensitive on the symmetry of the order parameter, thus it is a
useful tool to probe the pairing symmetry. A well-known result of
$d$-wave pairing in the cuprates is the zero bias peak in the local
density of states (LDOS) near the nonmagnetic impurity site. For the
case of iron-based materials, it was proposed that the impurity
scattering should be sensitive to the sign change of the
$s_{\pm}$-symmetry order parameter~\cite{cve}. Furthermore, the
averaged effect of an ensemble of impurities on the thermodynamic
properties has been studied in iron-based
materials~\cite{chub,bang,park,seng,voro}. The impurity-induced
intra-gap state forms due to the inter-band scattering~\cite{seng}
for the sign-reversing fully gapped order parameter symmetry. We
note that the calculation in Ref~\cite{seng} is based on a
phenomenological approach, namely, they treat the intra-band
scattering and inter-band scattering intensity as input parameters
and consider an impurity concentration. However, such intra-gap
state is not observed in scanning tunneling microscopy (STM)
experiments~\cite{ode,boy,panmh,yin}. In fact, the impurity-induced
intra-gap states are expected to exist only at or near the impurity
site. To study the possibility and property of such states one
effective method is to put one or several pointlike impurities in
real space and calculate the LDOS at or near the impurity site.
Basically two methods can be used, namely, the analytic $T$-matrix
approach and the self-consistent Bogoliubov-de Gennes (BdG)
equations. Based on the first method and taking into account the
$s_{\pm}$ symmetry phenomenologically, very recently
Ref.~\cite{zhang} calculated the LDOS near the impurity site and do
find the sharp resonance peaks inside the gap. The peaks will
disappear if the phase of the order parameter is taken to be the
same. The intra-gap bound states are actually located at the
positive and negative energies and the LDOS tends to be zero at low
energies. The impurity effect is expected to be sensitive to the
detailed band structure~\cite{bal}. The sensitivity maybe more
remarkable for the multi-band materials. Since the calculation in
Ref.~\cite{zhang} is based on a band structure which artificially
fits the ARPES experiments ~\cite{sek,tera}, in the present paper we
wish to examine the LDOS at and near the impurity site based on a
different two-band model as proposed by Raghu $et$ $al.$~\cite{rag},
and the self-consistent BdG equations. The advantage of the BdG
technique is that it calculates the order parameter in a fully
self-consistent way. The gap magnitude is expected to be suppressed
at and near the impurity site. It is proposed that such suppression
effect needs to be taken into account to study the local impurity
states~\cite{shni}, i.e., the suppression effect would enhance the
intensity of the bound states and the peak width and position also
depend on the amplitude of the suppression. For the case of the
$s_{x^2y^2}$ pairing symmetry, our numerical results show that the
intra-gap states exist as long as the scattering potential is strong
enough, but the detailed spectra of the local impurity states are
quite different and sensitive to the parameters for obtaining the
band structure.
 For a nonmagnetic impurity with positive scattering potential, we obtain two
intra-gap resonance peaks. The resonance peaks are located near the
coherent peaks for weak potentials and move to lower energy as the
potential increases. The weight of the resonance peak with positive
energy
 becomes very pronounced as the scattering potential reaching the unitary limit.
For the cases of negative scattering potentials,
 the intra-gap resonance peaks are moving away from the coherent
 peaks and are closer to the Fermi energy. The position of the resonance peaks are robust and
 depend weakly on the scattering potentials. All other features
 remain practically the same. We also studied the LDOS spectra for
 the cases of the sign unchanged $s$-wave and nodal $d_{x^2-y^2}$-wave
 and no intra-gap peaks are observed. Experimentally the LDOS can be measured by STM experiments,
 and thus is a useful tool to probe the pairing symmetry of this family of compounds.

The paper is organized as follows. In Sec. II, we introduce the
model and work out the formalism. In Sec. III, we persent the
numerical calculations and discuss the obtained results. Finally, we
give a brief summary in Sec. IV.

\section{Model and formalism}

 As we mentioned above, iron-based
superconducting materials have a layered structure. In the FeAs
layer there are two Fe ions per unit cell by taking into account As
ions located above or below the Fe-Fe plane. Currently no consensus
has been reached on what is the suitable model to capture the
essential physics~\cite{rag,kors,kuro,nakam,mala,palee,dagh,more}.
Some groups argue all five 3$d$ orbitals have to be considered to
construct a minimal model~\cite{kuro,nakam}. On the other hand, the
band calculations have also shown that the main features of the
bands that determine the Fermi surface are the two orbitals $d_{xz}$
and $d_{yz}$~\cite{boe,vild}. Thus the two band model is proposed by
several groups~\cite{rag,dagh,more}. Because we are concerned with
the low energy physics in the present work, we expect that the
qualitative result does not depend on the detailed band structure.
We here use the two band model considering the two degenerate
orbitals $d_{xz}$ and $d_yz$ per site~\cite{rag}. On the
two-dimensional square lattice, the model Hamiltonian reads,
\begin{equation}
H=H_t+H_\Delta+H_{imp},
\end{equation}
where $H_t$ is the hopping term,
\begin{equation}
H_t=-\sum_{i\mu j\nu\sigma}(t_{i\mu
j\nu}c^\dagger_{i\mu\sigma}c_{j\nu\sigma}+h.c.)-t_0\sum_{i\mu\sigma}c^{\dagger}_{i\mu\sigma}c_{i\mu\sigma}.
\end{equation}
where $i,j$ are the site indices and $\mu,\nu=1,2$ are the orbital
indices. $t_0$ is the chemical potential.

$H_\Delta$ is the pairing term,
\begin{equation}
H_\Delta=\sum_{i\mu j\nu\sigma}(\Delta_{i\mu
j\nu}c^\dagger_{i\mu\sigma}c^{\dagger}_{j\nu\bar{\sigma}}+h.c.).
\end{equation}

$H_{imp}$ is the impurity term, which is expressed by,
\begin{equation}
H_{imp}=\sum_{i_m\mu\nu\sigma}V_{s\mu\nu}c^{\dagger}_{i_m\mu\sigma}c_{i_m\nu\sigma},
\end{equation}
where $i_m$ represents for the impurity site and $V_{s\mu\nu}$ is
the scattering potential.

\begin{figure}
\centering
  \includegraphics[width=4in]{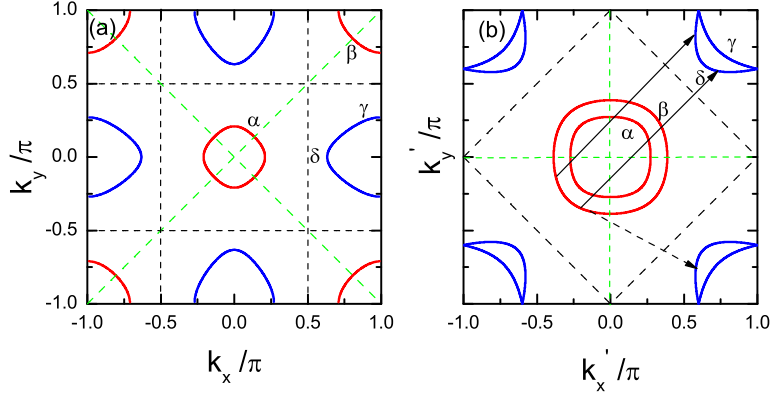}
\caption{The Fermi surface in the extended Brillouin zone and folded
Brillouin zone
  respectively. The black and green dashed lines are nodal lines for $s_{x^2y^2}$-wave, and $d_{x^2-y^2}$-wave pairing
  symmetry, respectively.
  The red and blue colors represent the hole and electron pockets, respectively.} \label{fig1}
\end{figure}

The Hamiltonian can be diagonalized by solving the BdG equations
self-consistently,
\begin{equation}
\sum_j\sum_\nu \left( \begin{array}{cc}
 H_{i\mu j\nu} & \Delta_{i\mu j\nu}  \\
 \Delta^{*}_{i\mu j\nu} & -H^{*}_{i\mu j\nu}
\end{array}
\right) \left( \begin{array}{c}
u^{n}_{j\nu\sigma}\\v^{n}_{j\nu\bar{\sigma}}
\end{array}
\right) =E_n \left( \begin{array}{c}
u^{n}_{i\mu\sigma}\\v^{n}_{i\mu\bar{\sigma}}
\end{array}
\right),
\end{equation}
where the Hamiltonian $H_{i\mu j\nu}$ is expressed by,
\begin{equation}
H_{i\mu j\nu}=-t_{i\mu
j\nu}+(V_s\delta_{i,i_m}-t_0)\delta_{ij}\delta_{\mu\nu}.
\end{equation}

The superconducting order parameter and the local electron density
$n_{i\mu}$ satisfy the following self-consistent conditions,
\begin{eqnarray}
\Delta_{i\mu j\nu}=\frac{V_{i\mu j\nu}}{4}\sum_n
(u^{n}_{i\mu\uparrow}v^{n*}_{j\nu\downarrow}+u^{n}_{j\nu\uparrow}v^{n*}_{i\mu\downarrow})\tanh
(\frac{E_n}{2K_B T}),
\end{eqnarray}
\begin{eqnarray}
n_{i\mu}&=&\sum_n |u^{n}_{i\mu\uparrow}|^{2}f(E_n)+\sum_n
|v^{n}_{i\mu\downarrow}|^{2}[1-f(E_n)].
\end{eqnarray}
Here $V_{i\mu j\nu}$ is the pairing strength and $f(x)$ is the Fermi
distribution function.

The LDOS is expressed by,
\begin{equation}
\rho_{i}(\omega)=\sum_{n\mu}[|u^{n}_{i\mu\sigma}|^{2}\delta(E_n-\omega)+
|v^{n}_{i\mu\bar{\sigma}}|^{2}\delta(E_n+\omega)],
\end{equation}
where the delta function $\delta(x)$ is taken as
$\Gamma/\pi(x^2+\Gamma^2)$, with $\Gamma=0.004$. The supercell
technical is used to calculated the LDOS.

In the following calculation, we use the hopping constant suggested
by Ref.~\cite{rag}, where $t_{1-3}$ are hoping constants between the
same orbital, expressed by,
\begin{eqnarray}
t_{i1,i\pm\hat{x}1}=t_{i2,i\pm\hat{y}2}=t_1=-1.0,\\
t_{i2,i\pm\hat{x}2}=t_{i1,i\pm\hat{y}1}=t_2=1.3 ,\\
t_{i\mu,i\pm\hat{x}\pm\hat{y}\mu}=t3=-0.85 \qquad  (\mu=1,2).
\end{eqnarray}
$t_4$ is the inter-orbital hopping constant, expressed by,
\begin{eqnarray}
t_{i\mu,i\pm(\hat{x}+\hat{y})\nu}=t_4=0.85 \qquad (\mu\neq \nu),\\
t_{i\mu,i\pm(\hat{x}-\hat{y})\nu}=-t_4=-0.85 \qquad (\mu\neq \nu).
\end{eqnarray}
$t_0$ is determined by the doping density. Throughout the work, the
energy is measured in units of $\vert t_1\vert = 1$. The Fermi
surface with filling electron density $2.1$ per site is plotted in
Fig.~\ref{fig1}(a). Alternatively, if we consider one unit cell
containing two Fe ions, the Fermi surface in the reduced Brillouin
zone is plotted in Fig.~\ref{fig1}(b).

The pairing symmetry is determined by the pairing potential $V_{i\mu
j\nu}$. We have carried out extensive calculations to search for
favorable pairing symmetry. To summarize, the on-site potential, the
nearest-neighbor (NN) potential, and the next-nearest-neighbor (NNN)
potential, will produce isotropic $s$-wave $(\Delta=\Delta_0)$,
$d_{x^2-y^2}$-wave $[\Delta=2\Delta_0 (\cos k_x-\cos k_y)]$, and
$s_{x^2y^2}$-wave $(\Delta=4\Delta_0 \cos k_x \cos k_y)$ pairing
symmetry, respectively. In the present work, we will focus on the
characteristic of $s_{\pm}$ symmetry. We consider the pairing
between the same orbital of the NNN site. Without the impurity,
self-consistent calculation verified that this kind of pairing
produces $s$-wave gap symmetry, i.e.,
$\Delta_{i\mu,i\pm\hat{x}\pm\hat{y}\mu}\equiv \Delta_0$.
Transforming the Hamiltonian to the momentum space and diagonalizing
the $4\times 4$ Hamiltonian, it is easy to verify that the pairing
symmetry is exactly the same with the $s_{x^2y^2}$-symmetry proposed
by Ref.~\cite{seo}, namely, the gap function in the extended
Brillouin zone has $ \cos k_x \cos k_y$ form, and has the form of
$\cos k_x^{\prime}+\cos k_y^{\prime}$ in the reduced Brillouin zone.
The nodal line of the gap function is shown in Fig. 1.

In the following presented results, we choose the filling electron
density $n=2.1$ per site (electron doped samples with doping
$\delta=0.1$) and pairing potential $V=1.2$. We consider the case
for the intra-orbital scattering, namely, $V_{s\mu
\nu}=V_s\delta_{\mu\nu}$. The parameters are chosen just for
illustration. We have checked numerically that our main results are
not sensitive to the parameters. The numerical calculation is
performed on $20\times 20$ lattice with the periodic boundary
conditions. The impurity is put at the site $(10,10)$. A $80\times
80$ supercell is taken to calculate the LDOS.

\begin{figure}
\centering
  \includegraphics[width=3.3in]{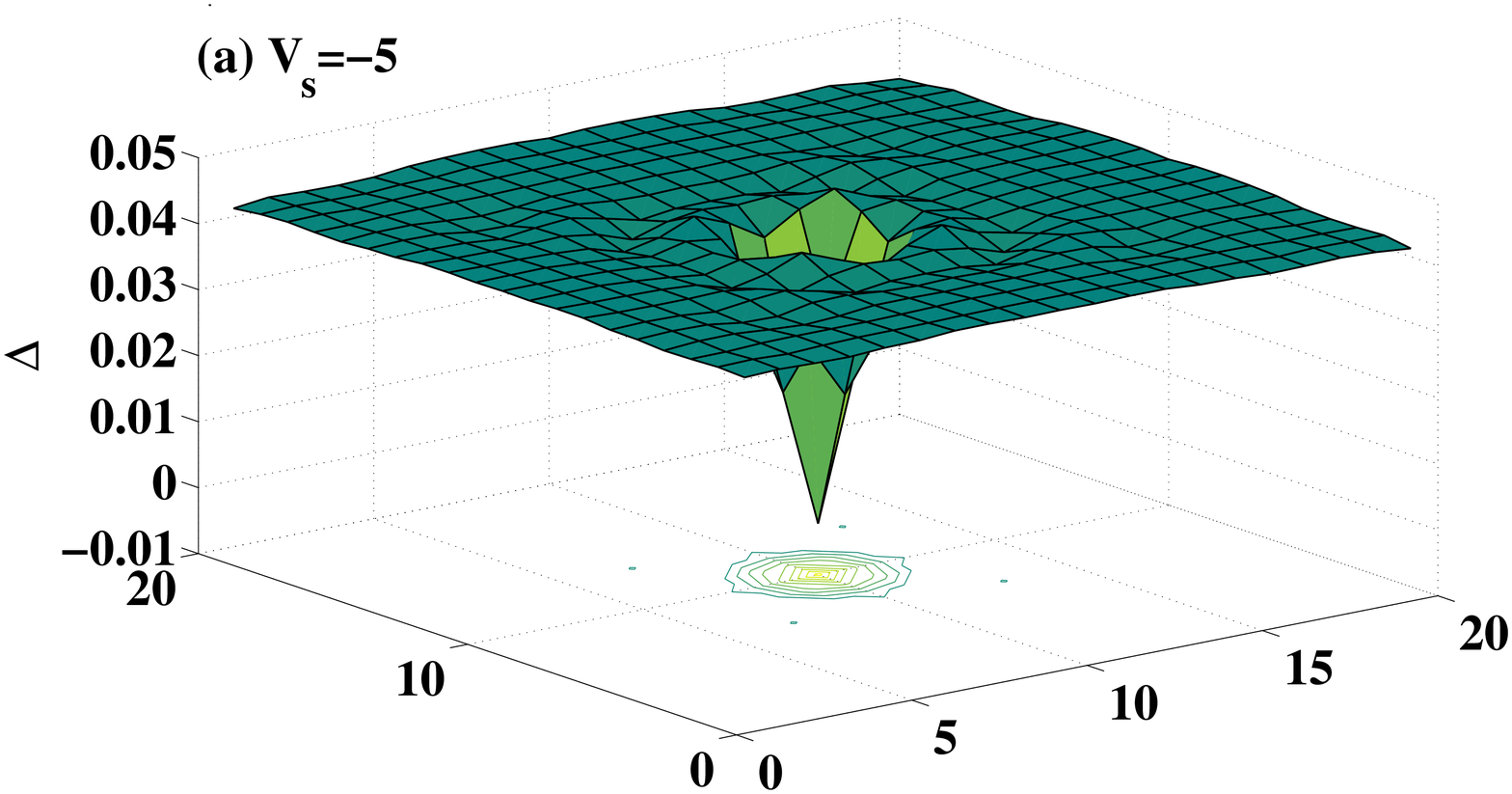}\\
 \vspace{6pt}
  \includegraphics[width=3.3in]{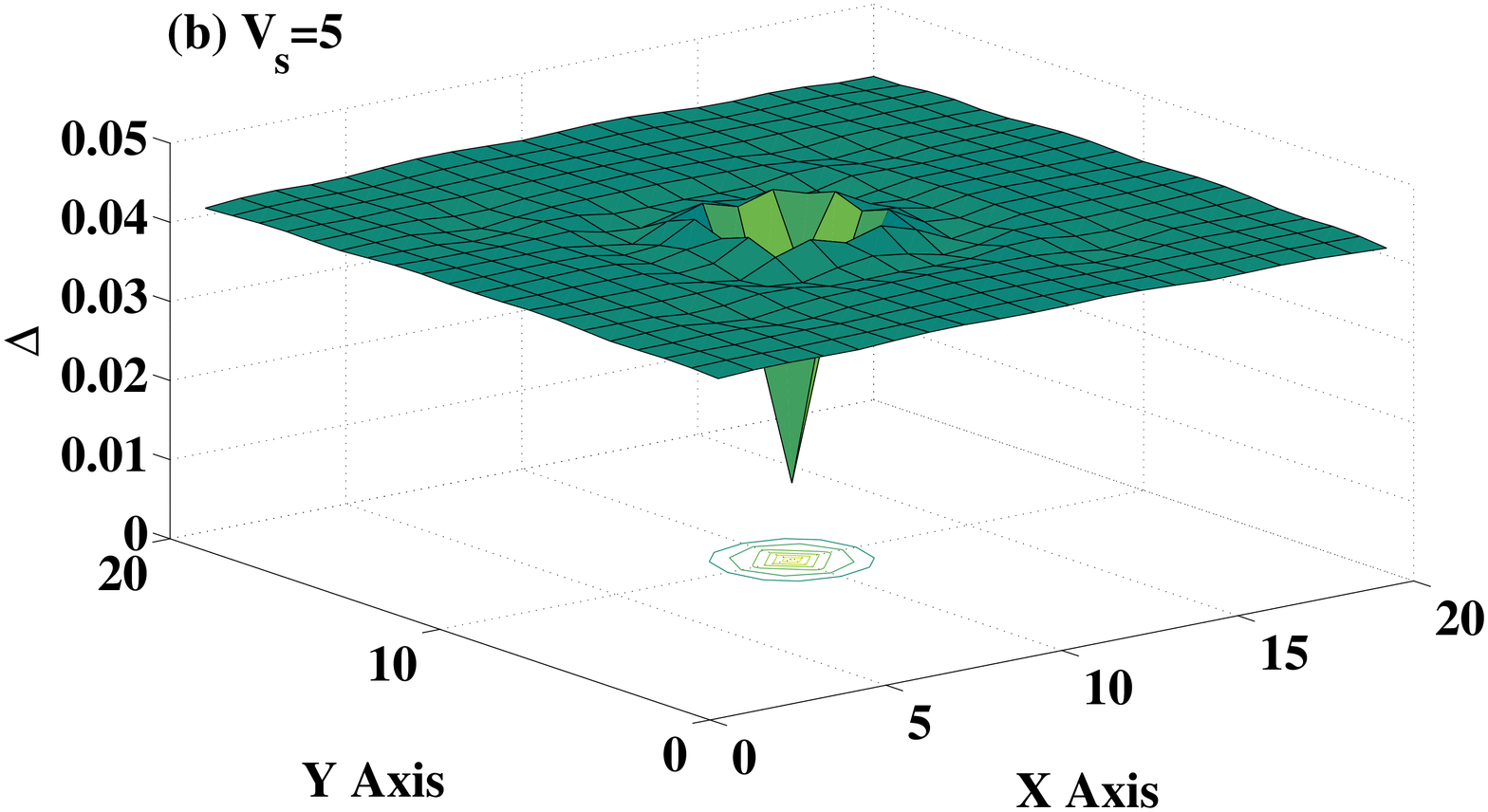}
\caption{Amplitudes of the order parameter in presence of a single
nonmagnetic impurity with $V_s=\mp 5$.} \label{fig2}
\end{figure}

\section{Results and discussion}

We first illustrate the feature of single nonmagnetic impurity
scattering with $s_{x^2y^2}$ pairing  symmetry. The scattering
potential $V_s$ can be estimated roughly by comparing the energy of
the impurity atoms and that of Fe atoms. In real materials, both
negative and positive scattering potentials are possible. Thus here
we discuss both cases. As it is unclear what an accurate value of
the potential strength is, we consider a range of potential values.

We plot the self-consistently determined order parameter amplitudes
on the lattice ($\Delta_{i}=\frac{1}{8}\sum_{j\mu}\Delta_{i\mu
j\mu}$) in Fig.~\ref{fig2} for $V_s=\pm5$. As seen, the order
parameter is suppressed near the impurity and quite small at the
impurity site. It will recover to the uniform value at about
$2\sim3$ lattice spacings from the impurity site. Note that the
suppression effect is more notable for the negative potential.

\begin{figure}
\centering
  \includegraphics[width=5in]{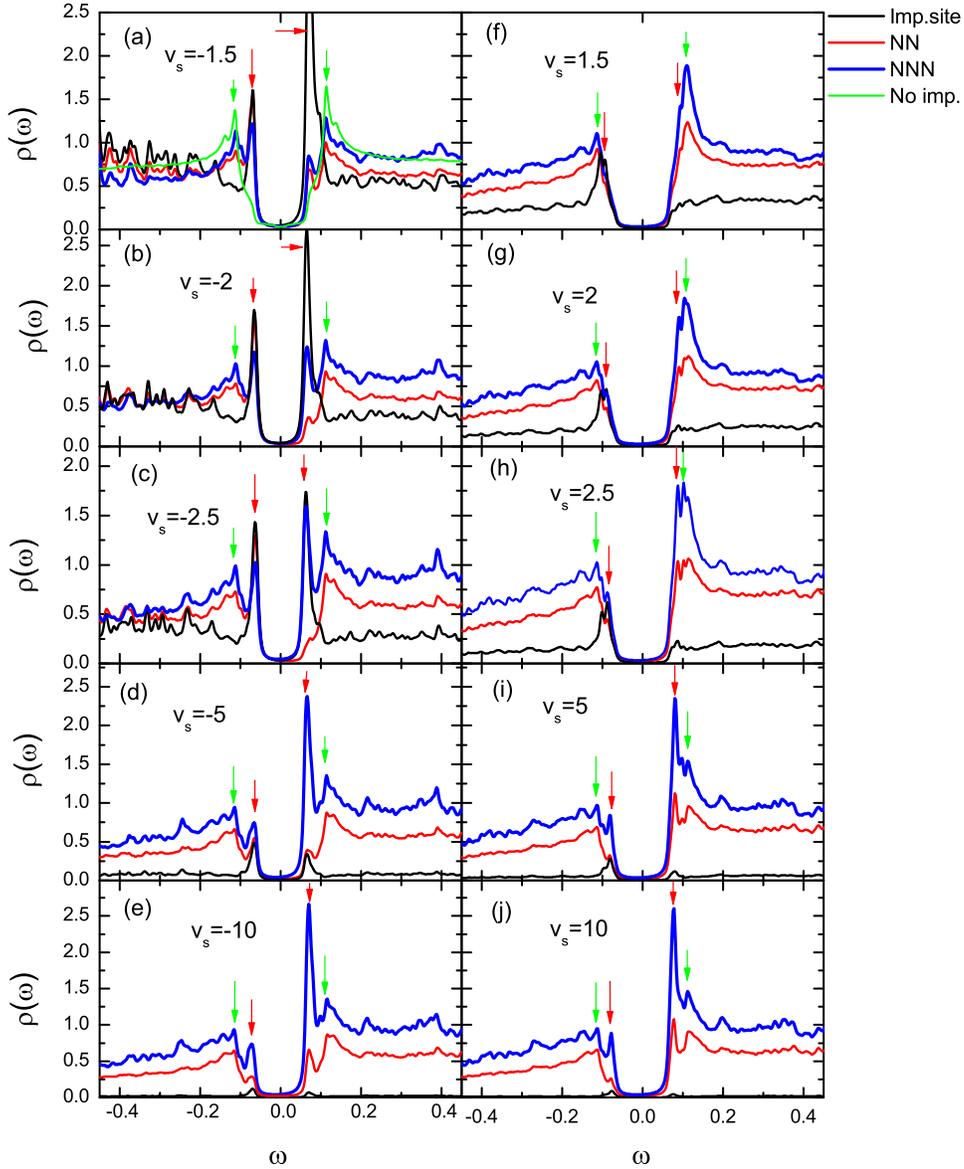}
\caption{The LDOS spectra in presence of a single nonmagnetic
impurity for different scattering potentials. The black lines are
the LDOS at the impurity site. The red and blue lines are the
spectra at NN and NNN sites to the impurity. The left and right
panels are for negative and positive scattering potential,
respectively. The green line in panel (a) is the density of states
spectrum without the impurity. The green and red arrows indicate the
superconducting coherent peak and intra-gap resonance peaks,
respectively.} \label{fig3}
\end{figure}

The LDOS spectra for different scattering potentials at and near the
impurity sites are plotted in Fig.~\ref{fig3}. The left panels are
the negative potential cases. Obvious impurity states (denoted by
the red arrows) exist at the energy about $\pm0.57 \Delta$ (here
$\Delta$ is the energy of the superconducting coherent peak, denoted
by the green arrows). The energy of the impurity resonance peaks
depends weakly on the scattering potential $V_s$. At the impurity
site, for small scattering potential $V_s$, two sharp resonances
peaks inside the gap are shown, and the resonance peaks are
suppressed as $V_s$ increases and the LDOS tends to zero at the
unitary limit. On the other hand, near the impurity site, the
intra-gap resonance peaks show up and the intensity increases as the
potential $V_s$ increases. The resonance peaks are remarkable at the
NNN site. In the unitary limit, a very sharp and strong resonance
peak appears at the positive energy for the LDOS at the NNN site of
the impurity site.  This situation may qualitatively correspond to
the case for a charged  Ba$^{+2}$ impurity with a negative or
attractive potential seen by electrons on the FeAs layer. This type
of impurity may be present in the FeAs layer when
BaFe$_{2-x}$Co$_{x}$As$_{2}$ is used for STM experiments~\cite{pan}.

The right panels show the effect of a positive scattering potential
impurity. We do not see clear resonance peaks at the impurity site
for the weak potential, which is different from that of the negative
potential. Near the impurity site, similar to the case of the
negative potential, the intra-gap resonance peaks show up as the
potential increases, while the intensities seem to be smaller than
those of the negative potential impurities. The resonance peaks are
more remarkable at the NNN site to the impurity, which is similar to
that of the negative potential. We also point out that
 the positions of the two resonance peaks are much closer to the
 coherent peaks for weak scattering potentials.
 The intensity of
the resonance peaks with positive energy increases and the positions
move away from the superconducting coherent peaks to lower energies
as the scattering potential increases. For a quite strong impurity
scattering potential ($V_s=10$), the intra-gap peaks are located at
the energy about $\pm 0.7 \Delta$. The spectrum is quite similar to
that of the negative potential $(V_s=-10)$.

\begin{figure}
\centering
  \includegraphics[width=5in]{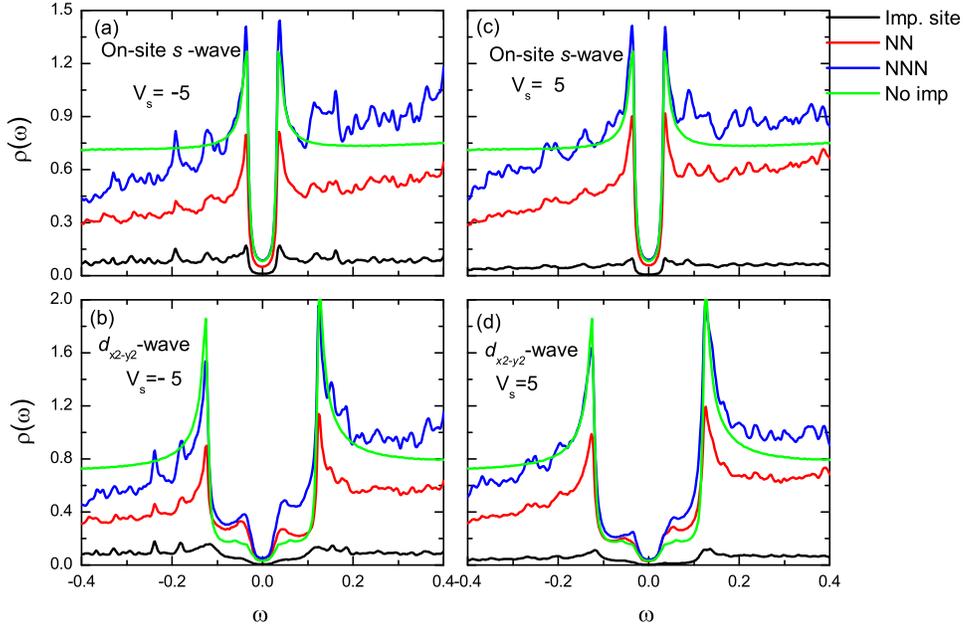}
\caption{The LDOS spectra in presence of a single nonmagnetic
impurity for on-site $s$-wave symmetry and $d_{x^2-y^2}$ symmetry
with $V_s=\pm5$, respectively. The black lines are the LDOS at the
impurity site. The red and blue lines are the spectra at NN and NNN
sites to the impurity. The green lines are density of states spectra
without the impurity.} \label{fig4}
\end{figure}

On the whole, we present the LDOS spectra taking into account the
impurity scattering for $s_{x^2y^2}$-pairing symmetry. The intra-gap
states for $s_{x^2y^2}$-pairing symmetry obtained here are quite
robust for a strong impurity. Experimentally, the intra-gap bound
states may be detected by comparing the LDOS spectrum far away the
impurity site with that near the impurity site. The intra-gap peaks
will show up near the impurity site. We also verified numerically
that the main results will not change for many impurities or the
impurity scattering potential is not only purely on-site but also
extends with a few lattice sites. While for quite weak scattering
potentials, the intensity of the intra-gap peaks would decrease and
disappear as the potentials are small enough. We have checked that
no obvious intra-gap features could be observed as $|V_s|<1$.

We now discuss the impurity scattering effect for different pairing
symmetries. The LDOS spectra at and near the impurity site for the
sign unchanged $s$-wave (on-site pairing) and $d_{x^2-y^2}$-wave (NN
pairing) with $V_s=\pm 5$ are shown in Fig.~\ref{fig4}. For the case
of the sign unchanged $s$-wave pairing symmetry, we can see clearly
two superconducting coherent peaks with the energies equal to those
of the density of states without the impurity. The intensity of the
coherent peaks are suppressed at the impurity site and its NN site.
No intra-gap peaks are observed. For the case of the $d$-wave
pairing symmetry, two intra-gap plateaus can be observed inside the
gap. The spectra are significantly different with those of the
cuprates. In fact, here the calculations are based on the two-band
model. For the $d_{x^2-y^2}$-wave order parameter, the hole pockets
around the $\Gamma$ point are nodal and the electron pockets around
the $M$ point are nodeless. As a result, the low energy spectra are
mainly contributed by the hole pockets, which lead to the intra-gap
plateaus. The interband scattering may enhance the intensity of the
plateaus while no intra-gap bound states are observed for
$d_{x^2-y^2}$ symmetry.

On the whole, we present the LDOS spectra taking into account the
impurity scattering for different pairing symmetries. The intra-gap
states for $s_{x^2y^2}$-pairing symmetry obtained here are quite
robust for a strong impurity. The electron operator for each orbital
is a linear combination of quasiparticle operators from all bands,
therefore the four Fermi surface pockets in Fig.1(b) are contributed
by the hybridigation of the two orbitals. With this observation, one
can find out that even the intra-orbital impurity scattering will
give rise to the scattering between different bands with the
scattering strength weighted by the Bogoliubov
amplitude~\cite{graser}. Comparing the LDOS spectra for the
$s_{x^2y^2}$ pairing symmetry shown in Fig.3 and those for the
sign-unchanged $s$-wave symmetry shown in Figs. 4(a) and 4(b), we
expect that the intra-gap bound state for the $s_{x^2y^2}$ pairing
symmetry is due to sign change of the order parameter between the
different Fermi surface pockets. As a result, we propose that this
can be used to detect the sign reversal of the order parameter.
However, we stress that the sign change of the order parameter is
just one of the essential conditions for the bound states but not
sufficient. The scattering is weighted by the Bogoliubov amplitude
so the band structure and the detailed scattering potential form is
also important. For the current model, the $d$-wave pairing symmetry
will not produce the bound states, as seen in Figs.4(c) and 4(d). We
here focus our discussion on the properties of the $s_{\pm}$ pairing
symmetry. The reason for the absence of the bound states for
$d$-wave pairing symmetry can be investigated further through the
momentum space scattering potential based on the $T$-matrix method
while this issue is not concerned here.

Comparing all the relevant results with different pairing
symmetries, we propose that one strong nonmagnetic impurity can be
used to probe the pairing symmetry and detect the sign reversal of
the order parameter.

\section{Summary}

In summary, we have studied the point-like single non-magnetic
impurity scattering effect in iron-based superconductors based on
self-consistent BdG equations. The LDOS at and near the impurity
site are calculated and the spectra have their unique features for
different pairing symmetry. Especially, for the case of the
$s_{x^2y^2}$-wave symmetry, two intra-gap peaks show up as the
impurity scattering potential is strong enough. This may also be
easily detected by the STM experiments through substituting the Fe
atom with a nonmagnetic atoms (e.g., Zn or Ba atoms etc.). As a
result, we propose that the non-magnetic impurity can be used to
probe the pairing symmetry of iron based superconductors.

$Note$ $added$- After we posted our work in Arxive, we note that
another group also did similar calculations~\cite{tsai} based on BdG
equations and T-matrix method using the same two band model proposed
by S. Raghu $et$ $al$. Their results are similar to ours.

\section*{acknowledgements}

The authors would like to thank S. H. Pan, Ang Li, and Degang Zhang
for useful discussions. One of us (X.H.) also acknowledges the
hospitality of Los Alamos National Laboratory (LANL), where part of
this work was initiated. This work was supported by the Texas Center
for Superconductivity at the University of Houston and by the Robert
A. Welch Foundation under the Grant no. E-1146 (T.Z., X.H., C.S.T.)
and the U.S. DOE at LANL under Contract No. DE-AC52-06NA25396, the
U.S. DOE Office of Science, and the LANL LDRD Program (J.X.Z.).


\section*{References}
\begin{thebibliography}{10}


\bibitem{kam} Kamihara Y $et$ $al.$ 2008 {\it J. Am. Chem. Soc.} {\bf 130} 3296
\bibitem{sin} Singh D J and Du M H 2008 {\it Phys. Rev. Lett.} {\bf 100} 237003
\bibitem{ish} Ishibashi S, Terakura K, and Hosono H 2008 {\it J. Phys. Soc. Jpn. {\bf 77}} 053709
\bibitem{nak} Nakamura K, Arita R, and Imada M 2008 {\it J. Phys. Soc. Jpn.} {\bf 77} 093711
\bibitem{lu} Lu D H $et$ $al.$ 2008 {\it Nature (London)} {\bf 455} 81
\bibitem{din} Ding H $et$ $al.$ 2008 {\it Europhys. Lett.} {\bf 83} 47001

\bibitem{kon} Kondo T $et$ $al.$ 2008 {\it Phys. Rev. Lett.} {\bf 101} 147003
\bibitem{naka} Nakayama K $et$ $al.$ 2009 {\it Europhys. Lett.} {\bf 85} 67002
\bibitem{evtu}  Evtushinsky D V $et$ $al.$ 2009 {\it Phys. Rev.} B {\bf 79} 054517
\bibitem{cliu} Liu C $et$ $al.$ 2008 {\it Phys. Rev. Lett.} {\bf 101} 177005
\bibitem{sad} For a review, see, e.g., Sadovskii M V 2008 {\it Phys. Usp.} {\bf 51} 1201
\bibitem{maz} Mazin I I, Singh D J, Johannes M D, and Du M H 2008 {\it Phys.
Rev. Lett.} {\bf 101} 057003
\bibitem{chu} Chubukov A V, Efremov D V, and Eremin I 2008 {\it Phys.
Rev.} B {\bf 78}, 134512
\bibitem{zjyao} Yao Z-J, Li J-X, and Wang Z D 2009 {\it New J. Phys.} {\bf 11}, 025009
\bibitem{shun} Yu S-L, Kang J, and Li J-X 2009 {\it Phys. Rev.} B {\bf 79} 064517
\bibitem{wan} Wang F, Zhai H, Ran Y, Vishwanath A, and Lee D-H 2009 {\it Phys.
Rev. Lett.} {\bf 102}, 047005
\bibitem{cve} Cvetkovic V and Tesanovic Z 2009 {\it Europhys. Lett.}
{\bf 85} 37002
\bibitem{has} Hashimoto K $et$ $al.$ 2009 {\it Phys. Rev. Lett.} {\bf 102} 017002

\bibitem{mu} Mu G, Luo H, Wang Z, Shan L, Ren C, Wen H-H 2009 {\it Phys. Rev.} B {\bf 79} 174501
\bibitem{tsa} Tsai W-F, Yao D-X, Bernevig B A, Hu J P 2009 {\it Phys. Rev.} B {\bf 80} 012511
\bibitem{gha} Ghaemi P, Wang F, Vishwanath A 2009 {\it Phys. Rev. Lett.} {\bf 102} 157002
\bibitem{par} Parker D and Mazin I 2008 arxiv: 0812.4416.
\bibitem{wu} Wu J and Phillips P 2009 {\it Phys. Rev.} B {\bf 79}
092502
\bibitem{zha} Zhang Y-Y, Fang C, Zhou X, Seo K, Tsai W-F,
Bernevig B A, and Hu J P 2009 {\it Phys. Rev.} B {\bf 80} 094528
\bibitem{bal} Balatsky A V, Vekhter I, and Zhu J-X 2006 {\it Rev.
Mod. Phys.} {\bf 78} 373

\bibitem{chub} Chubukov A V, Efremov D V, and Eremin I 2008
{\it Phys. Rev.} B {\bf 78} 134512
\bibitem{bang} Bang Y, Choi H Y, and Won H 2009 {\it Phys. Rev.} B {\bf 79} 054529
\bibitem{park} Parker D, Dolgov O V, Korshunov M M, Golubov A A, and
Mazin I I 2008 {\it Phys. Rev.} B {\bf 78} 134524
\bibitem{seng} Senga Y and Kontani H 2009 {\it New J. Phys.} {\bf 11} 035005

\bibitem{voro} Vorontsov A B, Vavilov M G, and Chubukov A V 2009 {\it Phys. Rev.} B {\bf 79}
140507(R)


\bibitem{ode} Millo O, Asulin I, Yuli O, Felner I, Ren Z-A, Shen X-L, Che G-C, and
Zhao Z-X 2008 {\it Phys. Rev.} B {\bf 78} 092505

\bibitem{boy} Boyer M C, Chatterjee K, Wise W D, Chen G F, Luo J L, Wang N L, and
Hudson E W 2008 arXiv:0806.4400


\bibitem{panmh}  Pan M H, He X B, Li G R, Wendelken J F, Jin R, Sefat A S, McGuire M A, Sales B C, Mandrus D, and
Plummer E W 2008 arxiv:0808.0895

\bibitem{yin} Yin Y, Zech M, Williams T L, Wang X F, Wu G, Chen X H, and
Hoffman J E 2009 {\it Phys. Rev. Lett.} {\bf 102} 097002




\bibitem{zhang} Zhang D G 2009 {\it Phys. Rev. Lett.} {\bf 103}
186402
\bibitem{sek} Sekiba Y $et$ $al.$ 2009 {\it New J. Phys.} {\bf 11} 025020
\bibitem{tera} Terashima K $et$ $al.$ 2009 {\it Proc. Natl. Acad. Sci.
U.S.A.} {\bf 106} 7330
\bibitem{rag} Raghu S, Qi X-L, Liu C-X, Scalapino D J, and
Zhang S-C 2008 {\it Phys. Rev.} B {\bf 77} 220503(R)
\bibitem{shni} Shnirman A, Adagideli I, Goldbart P M, and Yazdani A
1999 {\it Phys. Rev.} B {\bf 60} 7517


\bibitem{kors} Korshunov M M and Eremin I 2008 {\it Phys. Rev.} B {\bf 78} 140509(R)
\bibitem{kuro} Kuroki K, Onari S, Arita R, Usui H, Tanaka Y, Kontani H,
and Aoki H 2008 {\it Phys. Rev. Lett.} {\bf 101} 087004
\bibitem{nakam} Nakamura K, Arita R, and Imada M 2008 {\it J. Phys. Soc. Jpn.} {\bf 77} 093711


\bibitem{mala} Malaeb W $et$ $al$. 2008 {\it J. Phys. Soc. Jpn.} {\bf 77} 093714
\bibitem{palee} Lee P A and Wen X-G 2008 {\it Phys. Rev.} B {\bf 78} 144517

\bibitem{dagh} Daghofer M, Moreo A, Riera J A, Arrigoni E, Scalapino D J, and
Dagotto E 2008 {\it Phys. Rev. Lett.} {\bf 101} 237004
\bibitem{more} Moreo A, Daghofer M, Riera J A, and Dagotto E, {\it Phys. Rev.} B {\bf 79}, 134502
(2009)

\bibitem{boe} Boeri L, Dolgov O V, and Golubov A A 2008 {\it Phys. Rev. Lett.} {\bf 101} 026403
\bibitem{vild} Vildosola V, Pourovskii L, Arita R, Biermann S, and Georges A 2008 Phys. Rev. B {\bf 78} 064518


\bibitem{seo} Seo K, Bernevig B A, and Hu J P 2008 {\it Phys. Rev. Lett.} {\bf 101} 206404
\bibitem{pan} Pan S H $et$ $al.$, private communication
\bibitem{graser} Graser S, Maier T A, Hirschfeld P J, and Scalapino
D J 2009 {\it New. J. Phys.} {\bf 11} 025016
\bibitem{andrey} Chubukov A V, Eremin I, and
Korshunov M M 2009 {\it Phys. Rev. B} {\bf 79} 220501(R)
\bibitem{tsai} Tsai W-F, Zhang Y-Y, Fang C, and Hu J P 2009
{\it Phys. Rev} B {\bf 80} 064513

\end{thebibliography}
\end{document}